\documentclass[prb,showpacs,twocolumn,aps,superscriptaddress,a4paper,footinbib]{revtex4-1}
\usepackage{dcolumn,amssymb,amsmath,amsfonts,graphicx,latexsym,color,braket,amssymb,mathrsfs}

\begin{document}

\title{Coherent transport through a resonant level coupled to random-matrix leads}
\author{Xinxin Yang}
\affiliation{Department of Physics, Zhejiang Normal University, Jinhua 321004, People's Republic of China}
\author{Pei Wang}
\email{wangpei@zjnu.cn}
\affiliation{Department of Physics, Zhejiang Normal University, Jinhua 321004, People's Republic of China}
\date{\today}

\begin{abstract}
We study the transport through a resonant level coupled to two leads
with the latter being described by Wigner's random matrices.
By taking appropriate thermodynamic limit before taking the long time limit,
we obtain the stationary current as a function of voltage bias.
The I-V curve is similar to that of single impurity Anderson model.
On the other hand, the current matrix and initial density matrix in our model
look like random matrices in the eigenbasis of Hamiltonian.
They satisfy the description of eigenstate thermalization hypothesis (ETH) and nonequilibrium
steady state hypothesis (NESSH), respectively. A statistical formula
of current has been derived based on ETH and NESSH
({\it J. Stat. Mech.: Theo. Exp., 093105 (2017)}).
We check this formula in our model and find it to predict the stationary current to a
high precision. The shape of I-V curve is explained by the peak structure in
the characteristic function of NESSH, which is reminiscent of the transmission coefficient.
\end{abstract}

\maketitle

\section{\label{sec:level 1}Introduction}

According to statistical mechanics, the property of an equilibrium system
is given by averaging over an ensemble of microscopic states.
But during a measurement in laboratory, the microscopic state of system
is deterministic at every moment, whose evolution follows the classical or quantum laws.
Why the ensemble theory correctly predicts the measurement results
has been a controversial problem since the foundation of statistical mechanics.
It becomes even more sophiscated under the framework of quantum mechanics,
where ergodicity breaks down due to the linearity of Schr\"{o}dinger equation.
A breakthrough was made by Wigner~\cite{Wigner55} who proposed
a random-matrix model to explain the level statistics of heavy nuclei.
Following his approach, quantum chaos theory was developed~\cite{Wigner57,Wigner58,Dyson1962}.
Later on, the eigenstate thermalization hypothesis (ETH) was proposed~\cite{Deutsch91,Srednicki94}
to explain the relation between ensemble theory and microscopic dynamics.
ETH assumes that a laboratory system can be initialized
in an arbitrary generic quantum state. But the off-diagonal
elements of the initial density matrix
stop contributing to the expectation of observable after
a short relaxation time~\cite{Alessio2016}. Once if the energy fluctuation is subextensive,
the distribution of diagonal elements is also unimportant,
because the expectation of observable with respect to eigenstates
is a smooth function of energy. For an arbitrary initial state,
the expectation of observable quickly relaxes to its average over
a diagonal ensemble, and the latter is indeed equal to the average
over a microcanonical ensemble~\cite{Rigol2008}. It is the observable that thermalizes,
instead of the density matrix.

During the 1980s, the development of nanotechnology pushes forward
the studies on transport through mesoscopic structures~\cite{Umbach84,Webb85,Reed88}.
The movement of electrons in these devices is basically quantum.
Within the single-particle picture, Landauer~\cite{Landauer57,Landauer70} obtained a conductance formula,
and B\"{u}ttiker~\cite{Buttiker86} generalized it into the situation of multiple terminals.
At the beginning of 1990s, the Keldysh Green's functions were
widely employed to compute the conductance~\cite{Meir92, Meir93}.
Motivated by the actual needs for non-perturbative
calculation of conductance, researchers made efforts to derive
an ensemble description for the current-carrying nonequilibrium steady states (NESS) just like
the Gibbs ensemble for equilibrium states.
Hershfield~\cite{Hershfield93} obtained a density matrix of NESS by
starting from a Gibbs ensemble and then solving the Schr\"{o}dinger equation in the long-time limit.
The density matrix can also be obtained by maximizing
the entropy subject to appropriate constraints~\cite{Bokes03}.
It was later proved by Ness~\cite{Ness13} that these approaches all result
in a McLennan-Zubarev~\cite{McLennan59,Zubarev94} nonequilibrium ensemble,
i.e. a generalized Gibbs ensemble.

Starting from some initial state, a system
will thermalize or evolve into NESS after long enough time~\cite{footnote}.
Here we consider closed systems and treat leads as parts of the system
when discussing mesoscopic transport.
For a closed system, the quantum state remains pure for all the time.
ETH explains why the expectation of observable can be predicted
by an equilibrium ensemble after the system thermalizes.
Similarly, one may ask why the expectation
of observable (such as current)  is predicted by the nonequilibrium ensemble
after the system evolves into NESS. Recall that the nonequilibrium ensemble
is obtained by a real-time evolution from some initial equilibrium ensemble.
If we follow the logic of ETH and assume that the laboratory
system can be initialized in an arbitrary generic state,
the above question becomes why generic states
at the initial time give the same current as an equilibrium ensemble after infinitely long evolution.
In other words, why is the memory of initial state lost in the evolution to NESS?

Let us make it clear that this question cannot be answered in the
same way of explaining thermalization.
There is a big difference between NESS and thermalized state.
For the latter, one can assume that the expectation of observable
is equal to its average over the diagonal ensemble, i.e. an ensemble of eigenstates.
But for NESS, this assumption must be abandoned because it
rules out the possibility of nonzero current (the expectation of current with respect to diagonal ensemble is zero).
The diagonal ensemble is reached only if the system is finite in size with nonzero
level spacing, which causes the off-diagonal
elements of initial density matrix being averaged out. But a finite-size system never evolves into
NESS, because the initial imbalance will be removed
within a finite period. The knowledge about off-diagonal
elements of initial density matrix is necessary for predicting the current in NESS.

In an attempt to answer the above question,
an assumption is proposed for the density matrix of a generic state,
which is dubbed the nonequilibrium steady state hypothesis (NESSH)~\cite{Wang_2017}.
It states that, in the eigenbasis of Hamiltonian, the off-diagonal elements
of density matrix after coarse-graining become a smooth function
of energies. By combining the assumption for the density matrix
and ETH for the observable matrix, one derives a formula
for the expectation of observable in NESS, especially the current.
This formula clarifies that the observable depends only upon
a few statistical features of the initial density matrix but
being independent of its detail, and then makes progress in
explaining why different initial states predict the same value of observable
in NESS.

NESSH was verified numerically in a few models~\cite{yang2018nonequilibrium}, but the current
formula based on it has not been checked. To check the current formula,
we need a model in which ETH and NESSH stand,
at the same time, the steady current can be computed precisely.
For the latter purpose, the size of the system needs
to be large enough and the Schr\"{o}dinger or Heisenberg equation
must be solved until a time that is much larger than the current relaxation time.
These conditions are fulfilled in some integrable
models, e.g. the single-impurity Anderson model (SIAM) without electron-electron interaction.
However, to the best of our knowledge, in such models
that has been studied so far, neither ETH nor NESSH stand.
ETH usually stands in chaotic models in which the long-time evolution
and the large system's size are not accessible simultaneously.

To circumvent this problem, we propose a model similar to
SIAM but with two leads being replaced
by random matrices. After the replacement, the single-particle density matrix
satisfies NESSH, while the current matrix in the single-particle
eigenbasis satisfies ETH. It is worth mentioning that,
the application of random matrices in modeling the mesoscopic transport
has a long history, which was originally motivated by the
need to explain universal conductance fluctuations~\cite{RevModPhys.69.731}
and becomes a regular tool today~\cite{Datta1997,Pikulin_2012,lovas2019theory}.
But in previous studies, it is the scattering region which is
described by a random matrix, instead of the leads.
In these studies, the whole system does not display features
of ETH or NESSH, because the leads are overwhelmingly large compared to
the scattering region. For ETH and NESSH to stand, we model the leads
by random matrices. This can also be justified from the aspect
of experiments. The leads manufactured in mesoscopic experiments
indeed have irregular shapes, unavoidable impurities,
electron-electron interaction and electron-phonon interaction.
It is then reasonable to expect that the level statistics of leads
is similar to that of the Wigner's random matrix.

The rest of paper is organized as follows. We review the assumptions
and formula of ETH and NESSH in Sec.~\ref{sec:ETHNESSH}.
Especially, we explain how to derive the current formula.
Our model of random-matrix leads is introduced in Sec.~\ref{sec:level 2}
together with the current operator. In Sec.~\ref{sec:level 3}, it is made clear that
both ETH and NESSH stand in our model, and we explain how to obtain the characteristic
functions which are needed in the current formula.
In Sec.~\ref{sec:current}, the current is obtained by using the current formula.
A comparison with the ab-initio calculation is presented to check the current formula.
Sec.~\ref{sec:level 4} is a summary.

\section{\label{sec:ETHNESSH} ETH, NESSH and current formula}

In this section, we shortly review the formula of ETH and NESSH.
Our review is based on the references~[\onlinecite{Alessio2016}]
and~[\onlinecite{Wang_2017}].

Let us consider a system which is prepared in a generic quantum state
denoted by $\ket{\Psi_0}$. The density matrix is correspondingly $\hat \rho_0 = \ket{\Psi_0} \bra{\Psi_0}$.
Here a few words are needed for explaining which states are "generic".
Obviously, if we consider the equal-weight superposition of two eigenstates
with different energies, it neither thermalizes
nor evolves into NESS. But such a well-tuned many-body state is
hard if not impossible to realize in experiments. It is not generic. Due to the unavoidable
interaction between particles in a laboratory system,
the eigenstate usually involves huge number of natural bases.
And generally speaking, the prepared initial state is either an eigenstate (such as the ground
state) or a superposition of huge number of eigenstates.
The latter is called "generic" in this paper.

We use $\hat H$ to denote the Hamiltonian and the Greek
letters such as $\alpha$ and $\beta$ to denote the eigenstates with
the eigenenergies $E_\alpha$ and $E_\beta$, respectively.
The evolution of density matrix follows the quantum Liouville equation, reading
\begin{equation}
\hat \rho(t) = e^{-i \hat H t} \hat \rho_0 e^{i\hat H t}.
\end{equation}
Suppose that $\hat I$ is an observable operator. Its expectation value
at time $t$ is given by $I(t)=\textbf{Tr}\left[ \hat\rho(t) \hat I\right]$.

One is interested in the fate of $I(t)$ in the asymptotically long time.
An argument based on ETH shows that $I(t)$ must relax to a stationary
value which is defined as $I=\displaystyle\lim_{t\to\infty} I(t)$. And this
stationary value is equal to the long-time average of $I(t)$, which
can be expressed as
\begin{equation}\label{eq:steadyo}
\begin{split}
I = &  \lim_{T\to\infty} \frac{1}{T} \int_0^T dt e^{ i t(E_\alpha-E_\beta)} \sum_{\alpha,\beta}
\left[ \bra{\beta}\hat\rho_0\ket{\alpha}  \bra{\alpha}\hat I \ket{\beta} \right] \\ 
= & \sum_\alpha \left[ \bra{\alpha}\hat\rho_0\ket{\alpha}  \bra{\alpha}\hat I \ket{\alpha} \right].
\end{split}
\end{equation}
Eq.~\eqref{eq:steadyo} shows that the off-diagonal elements -
$\bra{\beta}\hat\rho_0\ket{\alpha}$ with $\alpha\neq\beta$
do not contribute to $I$. For obtaining this conclusion,
we have to assume $E_\alpha \neq E_\beta$ for arbitrary $\alpha \neq \beta$
and $\left|E_\alpha - E_\beta\right|$ has a nonzero lower bound.
This nondegeneracy condition is fulfilled in a generic system if
the system's size is finite. Usually, the symmetries must be broken for nondegeneracy,
which is considered as a prerequisite in the discussion of thermalization.

ETH assumes that the matrix of $\hat I$ in the eigenbasis of $\hat H$ can be written as
\begin{equation}
\begin{split}\label{eq:ETH}
I_{\alpha,\beta} &= \bra{\alpha}\hat I \ket{\beta} \\ & = I(\bar{E})
\delta_{\alpha,\beta} + D^{-1/2}(\bar{E})f_I(\bar{E},\omega)R^I_{\alpha\beta},
\end{split}
\end{equation}
where $\bar{E}=(E_\alpha+E_\beta)/2$ is the average energy,
$\omega=E_\alpha-E_\beta$ is the energy difference, and $D(\bar{E})$
is the density of eigenstates. $I(\bar{E})$ and $f_I(\bar{E},\omega)$ are
smooth functions. $R^I_{\alpha\beta}$
is a random number with zero mean and unit variance. The first and
second terms of Eq.~\eqref{eq:ETH} are for the diagonal and off-diagonal
elements, respectively. Crucially, Eq.~\eqref{eq:ETH} tells us that the diagonal elements
are a smooth function of $\bar{E}$. It is natural to assume that the energy fluctuation of the
initial state is subextensive, hence, $\rho_{\alpha,\alpha}= \bra{\alpha}\hat\rho_0\ket{\alpha}$
is significant only within a small energy shell centered at $E(\hat\rho_0)=
\textbf{Tr}\left[ \hat\rho_0 \hat H \right]$, i.e. the energy of initial state. Therefore, the sum of $\alpha$
in Eq.~\eqref{eq:steadyo} can be treated as a sum over this energy shell,
within which $I_{\alpha,\alpha}$ is approximately a constant due to the smoothness of $I(\bar{E})$.
By using the relation $\sum_\alpha \rho_{\alpha,\alpha} \equiv 1$, one then concludes
that $I$ is equal to the value of $I(\bar{E})$ at $\bar{E}=E(\hat\rho_0)$. Furthermore,
since $I_{\alpha,\alpha}$ is a constant, $I$ must be also equal to the average
of $I_{\alpha,\alpha}$ over this energy shell, which is just the average over the microcanonical
ensemble. The expectation of observable finally relaxes to its value in a thermal ensemble.

Now we turn to NESS. A typical system hosting NESS includes two leads which are connected
to each other through a central scattering regime - a model
that was frequently used in the study of mesoscopic transport.
We are still interested in the expectation of an observable in the long time limit.
We choose an alternative strategy to evaluate $I$. 
We start from $I(t)$ which is
\begin{equation}
\begin{split}\label{eq:Itexp}
I(t) = \sum_\alpha  \rho_{\alpha,\alpha} I_{\alpha,\alpha} 
+ \sum_{\alpha\neq\beta}e^{-i\omega t}  \rho_{\alpha,\beta} I_{\beta,\alpha}.
\end{split}
\end{equation}
The first term of Eq.~\eqref{eq:Itexp} comes from
the diagonal elements, being equal to what we obtain from Eq.~\eqref{eq:steadyo}.
From now on, we use $I_{eq} $ to denote $\sum_\alpha  \rho_{\alpha,\alpha} I_{\alpha,\alpha} $.
If the second term of Eq.~\eqref{eq:Itexp} decays to zero in the limit $t\to\infty$,
the observable thermalizes and we find $I=I_{eq}$.
But if the second term relaxes to a nonzero value, the
system evolves into NESS instead and we have $I\neq I_{eq}$.
We use $I_{ne}(t)$ to denote the second term,
Eq.~\eqref{eq:Itexp} is then rewritten as $I(t)=I_{eq}+I_{ne}(t)$.

NESSH assumes
that $\rho_{\alpha,\beta} = \bra{\alpha}\hat\rho_0\ket{\beta}$ for a generic
state can be expressed as
\begin{equation}
\begin{split}\label{eq:NESSH}
\rho_{\alpha,\beta} = p (\bar{E})
\delta_{\alpha,\beta} + D^{-3/2}(\bar{E})f(\bar{E},\omega)R_{\alpha\beta},
\end{split}
\end{equation}
where $p (\bar{E})$ is a smooth function. $f(\bar{E},\omega)$ is
called the dynamical characteristic function, which is
smooth almost everywhere except for a measure-zero set
in the $\bar{E}$-$\omega$ plane. As similar as $R^I_{\alpha\beta}$,
$R_{\alpha\beta}$ is a random number of zero mean and unit variance.
Again, the first and second terms of Eq.~\eqref{eq:NESSH} are for
the diagonal and off-diagonal elements, respectively. In other words,
the second term is only for $\alpha \neq \beta$.
By combining the assumptions~\eqref{eq:ETH} and~\eqref{eq:NESSH},
we obtain an expression of $I_{ne}(t)$ which reads
\begin{equation}
\begin{split}\label{eq:Itexpoff}
I_{ne}(t) = 
\sum_{\alpha\neq\beta}e^{-i\omega t}  \frac{f(\bar{E},\omega) f_I(\bar{E},-\omega)}{D^{2}(\bar{E})}
R_{\alpha\beta} R^I_{\beta\alpha}.
\end{split}
\end{equation}
Next we divide the $E_\alpha$-$E_\beta$ plane (or $\bar{E}$-$\omega$ plane) into
many tiny boxes, with each one still containing a great
number of $\left\{(\alpha,\beta)\right\}$.
The functions $f$, $f_I$ and $D$ can be
treated as constants within each box, because they are smooth functions
of energy and the boxes are small. On the other hand,
$R_{\alpha\beta}$ and $ R^I_{\beta\alpha}$ are random numbers
depending on the microscopic states $\alpha$ and $\beta$.
In each box, $R_{\alpha\beta} R^I_{\beta\alpha}$ distributes diversely.
And we define the correlation function $\mathcal{C}$ to be
\begin{equation}
\mathcal{C} 
= \frac{1}{M} \sum_{(\alpha , \beta) \in box}R_{\alpha\beta} R^I_{\beta\alpha},
\end{equation}
where the sum is over a box centered at $(E_\alpha,E_\beta)$ and
$M$ is the total number of $(\alpha , \beta)$ within this box. It is natural to assume that
$\mathcal{C}$ is a smooth function of $E_\alpha$ and $E_\beta$. $\mathcal{C}$
is indeed the mean of the random number $R_{\alpha\beta} R^I_{\beta\alpha}$.

As the system's size increases, the level spacing between neighbor
eigenenergies vanishes gradually, we can
then replace the sum in Eq.~\eqref{eq:Itexpoff} by integral: $\sum_\alpha \to \int dE_\alpha \ D(E_\alpha)$.
Here we exclude the presence of isolated energy levels and
assume that $D$ is a regular function of energy, which is usually true
in the study of transport.
Moreover, we can change variables in the integral by using $\int dE_\alpha dE_\beta
= \int d\bar{E} d\omega $. $I_{ne}(t)$ becomes
\begin{equation}
\begin{split}\label{eq:Itexpoff2}
I_{ne}(t) = \int d\bar{E} d\omega \ &
e^{-i\omega t} \frac{D(\bar{E}+\omega/2)D(\bar{E}-\omega/2)}{D^2(\bar{E})}  \\
& \times f(\bar{E},\omega) f_I(\bar{E},-\omega) \mathcal{C}(\bar{E},\omega).
\end{split}
\end{equation}

It is prepared to discuss the long time limit of $I_{ne}(t)$, denoted by $I_{ne}
=\displaystyle \lim_{t\to\infty} I_{ne}(t)$. The integral with
respect to $\omega$ is crucial. Obviously, once if the system's size
is finite,  $\left|\omega\right| = \left| E_\alpha-E_\beta\right|$
has a nonzero lower bound. That is to say that the
integrand in Eq.~\eqref{eq:Itexpoff2} vanishes for $\omega$
being close to zero. According to Riemann-Lebesgue lemma, once if
the integrand  has no singularity at $\omega=0$, an integral of the type
$\int d\omega e^{-i\omega t} $ must vanish as $t\to\infty$.
This is what we expect, since a finite system always thermalizes
and $I_{ne}$ must be zero.

To see the conditions under which $I_{ne}$ is nonzero,
we need to know how a NESS is realized.
Let us consider a system consisting of two leads which are connected
to a scattering regime. Once if the leads are finite in length, the system
always thermalizes as $t\to\infty$. If the initial
chemical potentials of two leads are different, the system will
evolve into a current-carrying quasi-steady state and stays there
for some time, before it finally thermalizes. If we increase the length
of leads, the lifetime of the quasi-steady state also increases.
As the length of leads becomes infinite (thermodynamic limit), the quasi-steady
state becomes a true steady state, i.e. NESS. Therefore, the thermodynamic
limit must be taken before $t\to\infty$, if we hope to obtain a NESS.

In the thermodynamic limit, the level spacing vanishes.
In Eq.~\eqref{eq:Itexpoff2}, this means that the lower
bound of $\left|\omega\right| = \left| E_\alpha-E_\beta\right|$ vanishes,
and the integrand can be singular at $\omega=0$.
The second assumption of NESSH states that, an initial state
evolves into NESS if its dynamical characteristic function
is singular at $\omega=0$ with the form
\begin{equation}\label{eq:frho}
f(\bar{E},\omega) = \frac{\rho(\bar{E},\omega)}{\left| \omega \right|},
\end{equation}
where $\rho(\bar{E},\omega)$ is continuous at $\omega=0$.
To make further progress, we need to discuss the symmetry of
$\rho(\bar{E},\omega)$ and $f_I(\bar{E},\omega)$. For simplicity,
we suppose that the Hamiltonian matrix is real, which is typical
in the absence of magnetic field. The eigenvectors $\ket{\alpha}$
or $\ket{\beta}$ are now real vectors. If the initial state $\ket{\Psi_0}$
is a real vector, the matrix elements $\rho_{\alpha,\beta}$ are all real.
And from the hermitianity of $\hat \rho_0$, we know $\rho_{\alpha,\beta}
=\rho_{\beta,\alpha}$, which requires that $\rho(\bar{E},\omega)=
\rho(\bar{E},-\omega)$ be an even function of $\omega$ according
to Eq.~\eqref{eq:NESSH}. Note that $R_{\alpha,\beta}=R_{\beta,\alpha}$ is real.
Furthermore, we consider $\hat I$ to be the current operator.
It is usually defined as the change rate of particle number in one lead.
It is well known that such an operator is purely imaginary, i.e.
$I_{\alpha,\beta}$ is a purely imaginary number. From the hermitianity
of $\hat I$, we obtain $I_{\alpha,\beta} = I^*_{\beta,\alpha}=-I_{\beta,\alpha}$,
which requires that $f_I(\bar{E},\omega)=-f_I(\bar{E},-\omega)$ be
purely imaginary and an odd function of $\omega$.
Note that $R_{\alpha,\beta}^I=R^I_{\beta,\alpha}$ is real.

Substituting Eq.~\eqref{eq:frho} into Eq.~\eqref{eq:Itexpoff2} and taking
the limit $t\to\infty$, we immediately find
\begin{equation}\label{eq:IE}
I_{ne}=i\pi \int d\bar{E} \ \rho(\bar{E},0) f_I(\bar{E},0^+) \mathcal{C}(\bar{E},0),
\end{equation}
where $f_I(\bar{E},0^+)=\displaystyle\lim_{\omega\to 0^+} f_I(\bar{E},\omega)$.
To obtain Eq.~\eqref{eq:IE}, we change the variable $\omega t\to x$ and use
the relation $\int_{-\infty}^\infty dx \ e^{ix}/x = i\pi$.
The functions $\rho$ and $\mathcal{C}$ are continuous at $\omega=0$,
but $f_I(\bar{E},\omega)$ is not. This is why $f_I(\bar{E},0^+)$ appears in Eq.~\eqref{eq:IE}.
At the same time, $I_{eq}$ for the current operator must be zero.
To evaluate $I_{eq}$, we start from a finite system in which
the eigenstates and eigenenergies are well-defined, and then take
the thermodynamic limit. But in a finite system, the expectation
of current operator with respect to eigenstates is zero,
i.e., the diagonal elements $I_{\alpha,\alpha}$ are all zero. We
then obtain $I_{eq}=0$ for arbitrary system's size, thereafter, its thermodynamic
limit must also be zero. The stationary current then becomes
\begin{equation}
I=I_{ne}.
\end{equation}

Eq.~\eqref{eq:IE} is the current formula of NESSH.
It tells us that most elements of initial density matrix
do not contribute to the current in NESS. First, $I$ is independent of
the diagonal elements. Second, among the off-diagonal elements,
those with finite $\left| E_\alpha-E_\beta\right|$ have no contribution,
since both $\rho(\bar{E},0)$ and $\mathcal{C}(\bar{E},0)$ depend only upon
$\rho_{\alpha,\beta}$ in the asymptotic limit $\left| E_\alpha-E_\beta\right| \to 0$.
And $I$ is indirectly connected to $\rho_{\alpha,\beta}$
with infinitesimal energy difference. $\rho(\bar{E},0)$
is obtained by averaging $\rho_{\alpha,\beta}$ over a small
energy box, being then independent of the distribution of $\rho_{\alpha,\beta}$
within the box. And $\mathcal{C}(\bar{E},0)$ is the correlation between
$\rho_{\alpha,\beta}$ and $I_{\alpha,\beta}$ within the box.
Eq.~\eqref{eq:IE} shows that the current in NESS
is determined by a few statistics of initial density matrix and current matrix.
This result is reminiscent of a thermalization process,
during which the memory of initial microscopic state is lost and finally the
system's properties are determined by a few macroscopic parameters.
But NESS must be distinguished from an equilibrium state.
A finite stationary current survives in NESS but not in thermal equilibrium.

\section{\label{sec:level 2}Model}

We propose a model to check the current formula~\eqref{eq:IE}.
Our purpose is to compare the stationary current obtained from Eq.~\eqref{eq:IE} with
that obtained from an ab-initio calculation.
The prerequisite of using Eq.~\eqref{eq:IE} is that ETH and NESSH stand.
On the other hand, an ab-initio calculation requires us to
solve the Schr\"{o}dinger or Heisenberg equation for $I(t)$.
The thermodynamic limit and $t\to\infty$ are then taken in turn.
In the ab-initio calculation, we need to know $I(t)$
at both large $t$ and large system's size.
In our model, the prerequisite of Eq.~\eqref{eq:IE} is fulfilled
and an ab-initio calculation is accessible.

Our model is composed of a resonant level and two leads.
Crucially, the Hamiltonians of the leads are random matrices.
We start from two independent random matrices $A_1$ and $A_2$, which
are for the left and right leads, respectively. The distribution of $A_i$ ($i=1,2$)
is that of a Gaussian orthogonal ensemble (GOE)~\cite{kravtsov2009random} with
the probability density $P\left({A}_i\right)\propto 
exp\left[ -\text{Tr}({A}^2_i)/2\sigma^2\right]$, where $\sigma$ is related to
the averaged level spacing in leads. We diagonalize $A_i$ and obtain
a series of eigenvalues $\epsilon_{ik}$. The Hamiltonians of leads
are then expressed in the eigenbasis as
\begin{equation}\label{Hi}
\hat H_i=\sum_{k}\epsilon_{ik} \hat c^{\dagger}_{ik} \hat c_{ik},
\end{equation}
where $\epsilon_{ik}$ denotes the energy
levels of leads and $\hat c_{ik}$ and $\hat c^\dag_{ik}$ are the fermionic field operators.
$\epsilon_{ik}$ satisfies the well-known Wigner-Dyson distribution with
the probability density~\cite{kravtsov2009random}
\begin{equation}
P(\epsilon_{i1},\epsilon_{i2},\cdots,\epsilon_{i N_0}) \propto
e^{-\frac{\epsilon_{i1}^2+ \cdots + \epsilon_{iN_0}^2}{2\sigma^2}}
\left| \prod_{k'>k} \left(\epsilon_{ik'}-\epsilon_{ik}\right) \right|.
\end{equation}
$N_0$ denotes the dimension of $A_i$, which can be treated as the lead's size.
The thermodynamic limit is defined to be $N_0\to\infty$.
The averaged level spacing in leads is proportional to $\sigma$ according to
random matrix theory. The proper way of taking thermodynamic limit
is to increase the lead's size while keeping its bandwidth invariant.
Therefore, we keep $N_0\sigma$ a constant as $N_0$ increases.
In the numerical simulation, large level spacing
appears at the spectrum edge of $A_i$, which causes the density of
states ill-defined. To circumvent this problem, we keep only $\epsilon_{ik}$ that
lies within an interval $[-\Delta,\Delta]$ where $2\Delta$
denotes the lead's bandwidth. $\Delta$ should be small
compared to the largest eigenvalue of $A_i$, so that $\epsilon_{ik}$
distributes densely within the interval $[-\Delta,\Delta]$.

The Hamiltonian of the resonant level is $\hat H_d= \epsilon_d \hat d^{\dagger}\hat d$.
The coupling between the leads and the resonant level is set to
a random number. The corresponding Hamiltonian
is $\hat H_c=\sum_{i,k}g_{ik}\left(\hat c^{\dagger}_{ik}\hat d+H.c.\right)$. Here
$g_{ik}$ is an independent random number which has a Gaussian distribution with
the probability density $P({g_{ik}})=exp\left(-\left({g_{ik}}\right)^2/2\sigma^2_t\right)/\sqrt{2\pi\sigma^2_t}$,
where $\sigma_t$ denotes the coupling strength.
The total Hamiltonian is written as
\begin{equation}\label{toy}
\hat H= \hat H_1+\hat H_2 +\hat H_d +\hat H_c.
\end{equation}
We set $\epsilon_{jk}$ and $g_{ik}$ to be independent random numbers for ETH and NESSH to hold.
Otherwise, if the system has an explicit parity symmetry,
i.e. $g_{1k}=g_{2k}$ and $\epsilon_{1k}=\epsilon_{2k}$,
$\hat H$ can be decoupled by defining the
(anti)symmetric basis $\hat c_{k\pm}= \left(\hat c_{1k}\pm \hat c_{2k}\right) / \sqrt{2}$,
and then ETH and NESSH break down.
By using random levels and random couplings,
we keep the parity symmetry in a statistical sense.
Furthermore, for a resonant level coupled to leads, one usually defines
the broadening of resonant level as
$\Gamma(E) =\sum_{ik} \pi \delta(E-\epsilon_{ik}) g_{ik}^2$. 
The mean of $g_{ik}^2$ is defined to be $\sigma_t^2$.
It is then straight forward to obtain $\Gamma = 2\pi D_l \sigma_t^2$,
where $D_l$ is the density of states in one lead.
To take the proper thermodynamic limit, we must
keep $\Gamma$ invariant as the lead's size increases.
Therefore, $\sqrt{D_l} \sigma_t$ must be a constant as $N_0$ increases.

Our model excludes interaction between particles, otherwise,
an ab-initio calculation is hard to carry out.
In the absence of interaction, ETH and NESSH do not stand
for the many-body eigenstates. But they do stand under the single-particle
picture. If we use $\hat \rho$ to denote the density matrix
of a single particle and $\alpha$ and $\beta$ to denote the single-particle eigenstates, $\rho_{\alpha,\beta}$
and $I_{\alpha,\beta}$ satisfy Eq.~\eqref{eq:NESSH} and Eq.~\eqref{eq:ETH},
respectively. At the same time, due to the lack of interaction,
the ab-initio calculation of current is easy once if we suppose the initial state
to be a product of single-particle states.

In a two-lead model, the current operator $\hat I$ is usually defined as
the changing rate of particle number, reading
\begin{equation}\label{operator}
\begin{split}
\hat {I}=& \frac{1}{2}\left(\frac{d{\hat N}_2}{d t} -\frac{d{\hat N}_1}{d t} \right) \\ = & \frac{i}{2}\sum_{jk} (-1)^j
g_{jk}\left(\hat c^{\dagger}_{jk} \hat d- \hat d^{\dagger}\hat c_{jk}\right),
\end{split}
\end{equation}
where $\hat N_j=\sum_k \hat c^{\dagger}_{jk} \hat c_{jk}$ is the number of particles in lead $j$,
and $i$ is the imaginary unit. For obtaining a current-carrying NESS,
we employ next initial state. At time $t=0$, the levels between
$-V/2$ and $V/2$ in lead $1$ are occupied, while all the other levels
of lead $1$ or lead $2$ are empty.
The initial state is a product of left-lead occupied levels.
If we are only interested in the stationary current but not how $I(t)$ relaxes,
such defined initial state is equivalent to the initial condition under which
the left and right leads are at zero temperature with Fermi energies $V/2$
and $-V/2$, respectively. The latter condition is what one usually adopts in the
study of mesoscopic transport. $V$ has the meaning of voltage bias.
The equivalence between two initial conditions is due to the fact that
the current contributed by the levels lower than
$-V/2$ in lead $1$ neutralizes the current contributed by
the occupied levels in lead $2$ as $t\to\infty$.

We use $\hat \rho^{jk}$ to denote the density matrix
of a particle occupying the level $\epsilon_{jk}$ in lead $j$.
In the absence of particle-particle interaction, the total current can be expressed as
\begin{equation}\label{eq:Itsum}
I(t)=\sum_{jk} I^{jk} (t),
\end{equation}
where $I^{jk} (t)=\textbf{Tr} \left[ \hat I \hat\rho^{jk}(t) \right]$ is the current
contributed by a single particle and the sum
is over all the occupied levels at $t=0$.
By inserting the single-particle eigenstates $\alpha$ and $\beta$
into $\textbf{Tr} \left[ \hat I \hat\rho^{jk}(t) \right]$, Eq.~\eqref{eq:Itsum}
can be reexpressed as
\begin{equation}\label{t-current}
{I(t)}=\sum_{\alpha\ne\beta}e^{-i\left(E_{\alpha}
-E_{\beta}\right)t}\rho_{\alpha,\beta}I_{\beta,\alpha},
\end{equation}
where $\rho_{\alpha,\beta}=\sum_{j,k} \rho^{jk}_{\alpha,\beta}$.
And $\rho^{jk}_{\alpha,\beta}=\bra{\alpha} \hat  \rho^{jk} \ket{\beta}$
and $I_{\beta,\alpha}=\bra{\beta}{\hat I}\ket{\alpha}$ are
the single-particle density matrix and current matrix, respectively.
According to the definition of current operator,
it is easy to see $I_{\alpha\alpha}=0$. This explains why the terms with
$\alpha=\beta$ are excluded in Eq.~\eqref{t-current}.
Eq.~\eqref{t-current} has the same form as Eq.~\eqref{eq:Itexp},
except that $\rho_{\alpha,\beta}$ in Eq.~\eqref{t-current} is a sum of single-particle
density matrices instead of a many-body density matrix.
We find that such defined $\rho_{\alpha,\beta}$ satisfies Eq.~\eqref{eq:NESSH},
therefore, the current formula~\eqref{eq:IE} should stand in our model.
To check the current formula is to compare
Eq.~\eqref{eq:IE} with the current obtained from Eq.~\eqref{t-current}
by taking $t\to\infty$ after $N_0\to\infty$.

\begin{figure}[htbp]
	\centering
	\includegraphics[width=0.45\textwidth]{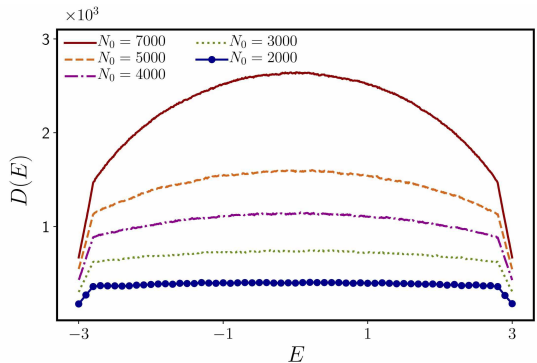}
	\caption{(Color online) The density of states for different $N_0$.
	We set $\epsilon_d=0$, $\Delta = 3$, $N_0\sigma=200$ and
	$\sqrt{D_l}\sigma_t = \sqrt{2/5}$.}\label{fig1}
\end{figure}
It is worth mentioning that sampling $g_{jk}$ or $A_j$ for multiple times
is unnecessary even they are random numbers or matrices. Once if $N_0$
is large enough, NESSH and ETH stand in each shot of sampling,
and the current formula can be checked by a single sampling.
Moreover, sampling for multiple times has little effect on the results of current.
We then do not carry out the sampling average in this paper.

\section{\label{sec:level 3} Evidence of ETH and NESSH}

In this section, we show that ETH and NESSH stand in our model.
And we compute the functions $D({E})$, $f_I(\bar{E},\omega)$,
$f(\bar{E},\omega)$ and $\mathcal{C}(\bar{E},\omega)$, which
are the characteristic functions of our model.
These functions are necessary in the application of current formula~\eqref{eq:IE}.

The free parameters of our model include $N_0$, $\Delta$, $N_0\sigma$,
$\sqrt{D_l} \sigma_t$ and $\epsilon_d$. We assign to each parameter a value.
The numerical simulation starts from generating the random matrices $A_i$
with respect to given $N_0$ and $\sigma$.
By diagonalizing $A_i$ we obtain all
the levels within lead $1$ and $2$. $D_l$ is the averaged density of
states in lead $1$, which is also equal to the averaged density of states in lead $2$.
We first compute $D_l(E)$ at a specific energy $E$
by counting the number of levels within a box of length $0.4$ centered at $E$.
We then obtain $D_l$ from $D_l =\frac{1}{2\Delta} \displaystyle \int_{-\Delta}^{\Delta} dE
D_l(E) $. Since $\sqrt{D_l} \sigma_t$ is a predefined parameter,
the value of $\sigma_t$ is obtained. Note that $g_{ik}$ is an
independent random number with normal distribution of variance $\sigma_t^2$.
We can now generate $g_{ik}$ and then the Hamiltonian matrix~\eqref{toy} in the single-particle basis.

Diagonalizing this matrix, we obtain all the eigenvectors and eigenenergies.
The density of states $D(E)$ is obtained by choosing an energy
box of length $0.4$ centered at $E$
and then counting the eigenenergies falling within it.
Here the size of box is chosen so that each box contains
a few hundreds of levels.
Fig.~\ref{fig1}  plots $D(E)$ for different $N_0$.
We see that $D(E)$ changes smoothly and
reaches a maximum at $E=0$. And the density of states increases
with the lead's size, as we expect.

\begin{figure}[htbp]
	\centering
	\includegraphics[width=0.5\textwidth]{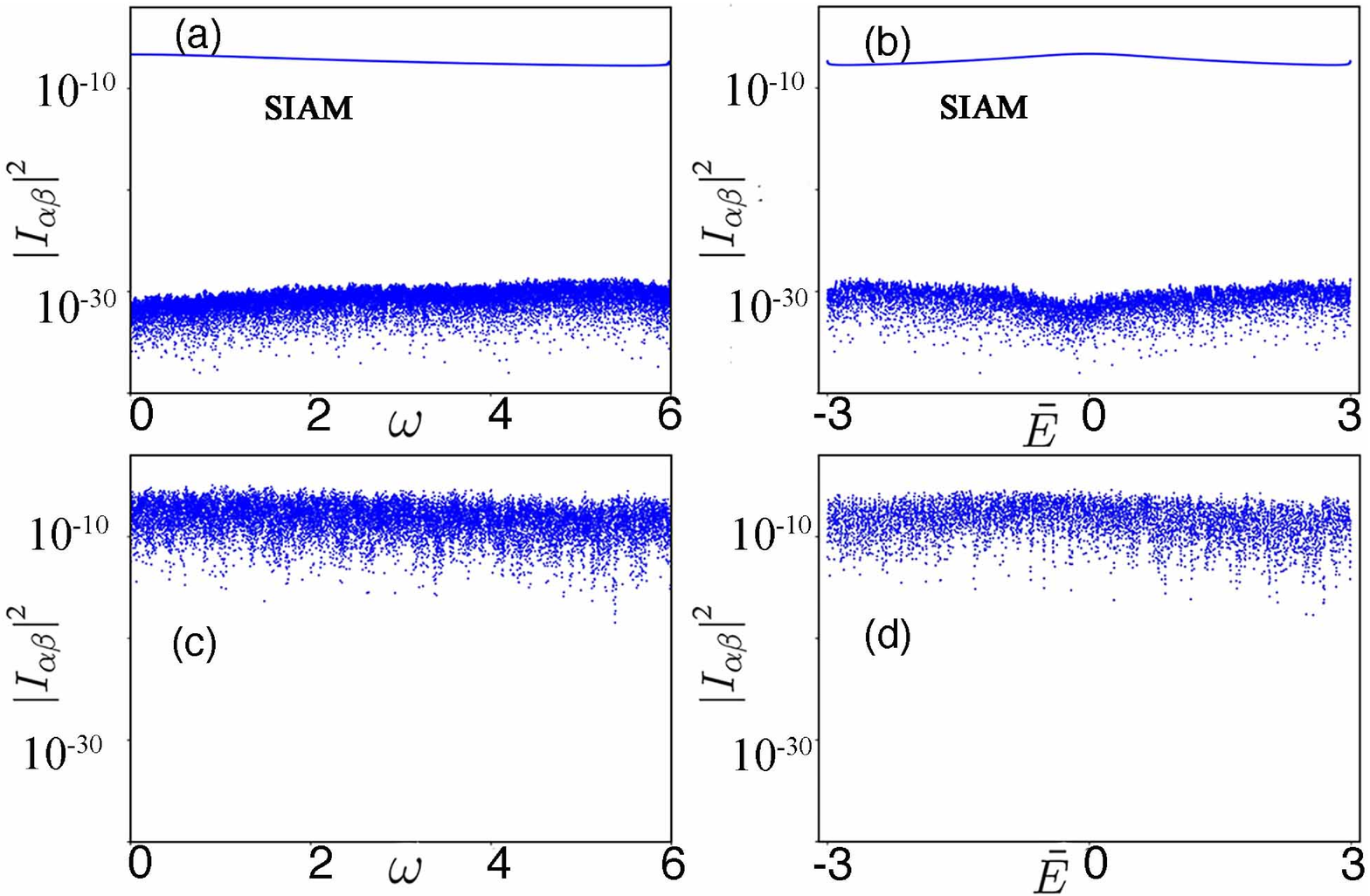}
	\caption{(Color online) (a) $\left| I_{\alpha,\beta}\right|^2$ vs $\omega$
	for a small window $\bar{E}\in(-0.01,0.01)$ in the SIAM.
	(b) $\left| I_{\alpha,\beta}\right|^2$ vs $\bar{E}$ for a small window
	$\omega\in(0,0.01)$ in the SIAM. (c) $\left| I_{\alpha,\beta}\right|^2$ vs $\omega$
	for a small window $\bar{E}\in(-0.01,0.01)$ in our model.
	(d) $\left| I_{\alpha,\beta}\right|^2$ vs $\bar{E}$ for a small window
	$\omega\in(0,0.01)$ in our model. The parameters of SIAM can
	be found in Appendix~\ref{app:SIAM}. For our model~(\ref{toy}),
	we set $\Delta = 3$, $\epsilon_d=0$, $N_o\sigma=200$, $N_0=2000$ and 
	$\sqrt{D_l} \sigma_t =1/\sqrt{10}$.}\label{fig1a}
\end{figure}
After we obtain all the eigenvectors, the current matrix can be computed
according to $I_{\alpha,\beta}=\bra{\alpha}\hat I\ket{\beta}$. Notice that
$I_{\alpha,\beta}$ is a purely imaginary number.
If one observes the distribution of $I_{\alpha,\beta}$,
it becomes clear that ETH stands in our model. The diagonal
elements of the current matrix are all zero, indicating $I(\bar{E})\equiv 0$ in Eq.~\eqref{eq:ETH}.
While Fig.~\ref{fig1a} plots the off-diagonal elements.
To be specific, Fig.~\ref{fig1a} plots $\left| I_{\alpha,\beta}\right|^2$ vs $\omega=E_\alpha-E_\beta$
for a small window $\bar{E}=(E_\alpha+E_\beta)/2\in (-0.01,0.01)$, and also
$\left| I_{\alpha,\beta}\right|^2$ vs $\bar{E}$ for a window $\omega \in (0,0.01)$.
As a comparison, $\left| I_{\alpha,\beta}\right|^2$ in the SIAM
is plotted in the same figure. The definition of SIAM
can be found in Appendix~\ref{app:SIAM}. The off-diagonal elements of current matrix
are qualitatively different in SIAM and our model. In the SIAM,
a large fraction of off-diagonal elements vanish
($\left| I_{\alpha,\beta}\right|^2 \sim 10^{-30}$), at
the same time, a small fraction of elements are significant
($\left| I_{\alpha,\beta}\right|^2\sim 10^{-10}$).
Conversely, we do not find large outliers in the off-diagonal elements of our model. 

If Eq.~\eqref{eq:ETH} stands, one must have $\left| I_{\alpha,\beta}\right|^2 
= D^{-1} \left| f_I\right|^2 \left| R^I_{\alpha\beta}\right|^2$, where $D$ and $f_I$ are
smooth functions of $\bar{E}$ and $\omega$ and $\left| R^I_{\alpha\beta}\right|^2$
is a random number of unit mean. As a consequence, $\left| I_{\alpha,\beta}\right|^2 $
should distribute densely around a smooth curve, i.e. $D^{-1}\left| f_I\right|^2$.
Obviously, $I_{\alpha,\beta}$ in SIAM cannot be described by Eq.~\eqref{eq:ETH}.
This is a result of SIAM's integrability. $I_{\alpha,\beta}$ in SIAM is significant
only for few $(\alpha,\beta)$ which are connected to the symmetry of the model.
On the other hand, these symmetries are broken in our model, thereafter,
$I_{\alpha,\beta}$ in our model can be described by Eq.~\eqref{eq:ETH},
as clearly shown in Fig.~\ref{fig1a}.

The function $f_I(\bar{E},\omega)$ is obtained by averaging $\left| I_{\alpha,\beta}\right|^2$
over an energy box. It is straight forward to see
\begin{equation}\label{eq:fImethod}
\left| f_I(\bar{E},\omega) \right|^2 = \overline{\left| I_{\alpha\beta}\right|^2} D(\bar{E}),
\end{equation}
where $\overline{\left| I_{\alpha\beta}\right|^2}$ denotes the average over
a square box centered at $(\bar{E},\omega)$ with the edge chosen to $0.2$.
Such a box usually contains thousands of $(\alpha,\beta)$. Eq.~\eqref{eq:fImethod}
does not tell us the sign of $f_I$, which can be decided freely. In the derivation
of Eq.~\eqref{eq:IE}, we have shown that $f_I$ is purely imaginary and an odd function
of $\omega$. Once if these properties are guaranteed, the sign of $f_I$ is unimportant,
because it can always be absorbed into $R^I_{\alpha,\beta}$.

\begin{figure}[htbp]
	\centering
	\includegraphics[width=0.45\textwidth]{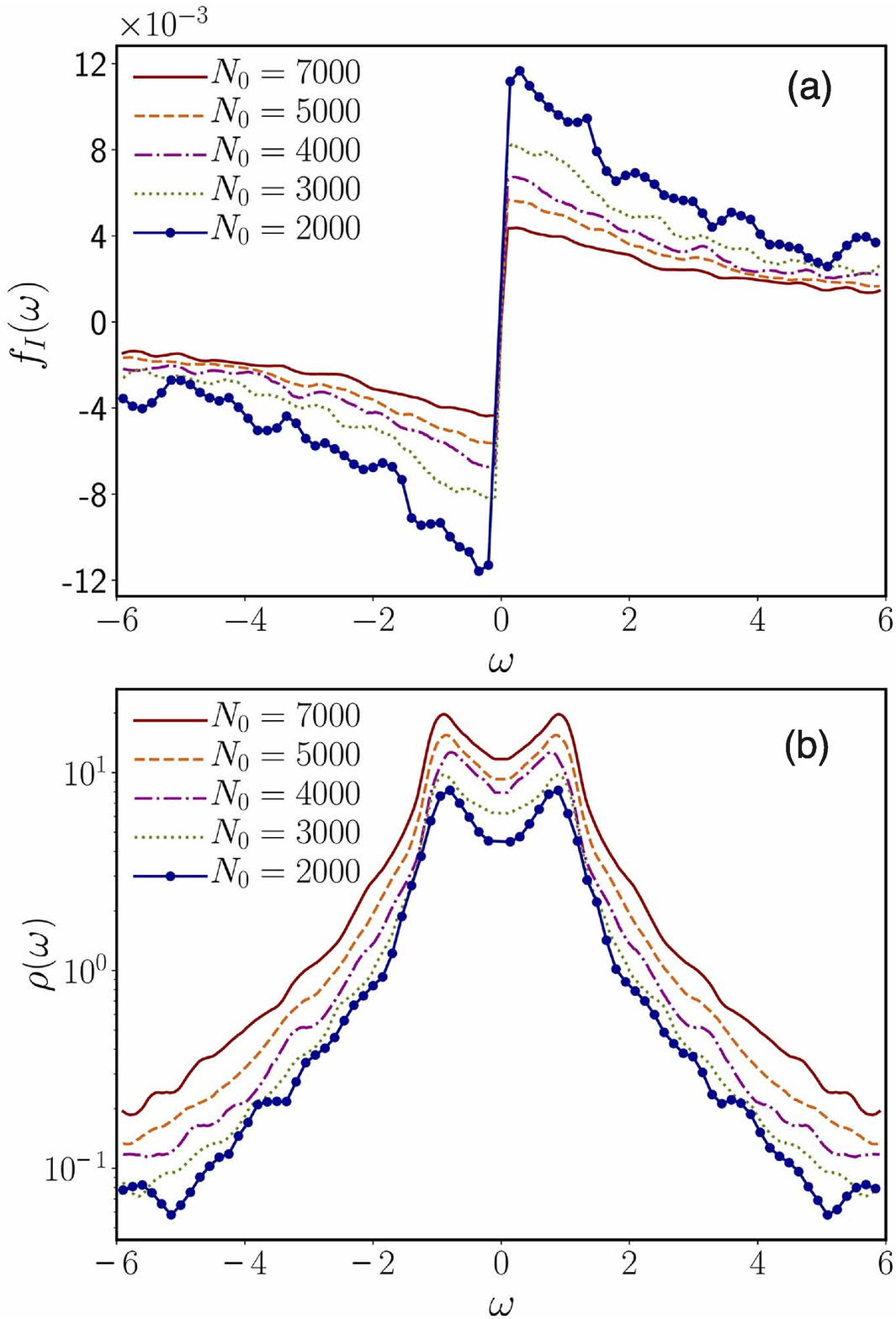}
	\caption{(Color online) (a) The function $f_I(0,\omega)$ for different $N_0$.
	(b) The function $\rho(0,\omega)$ for different $N_0$.
	The parameters are chosen to $\Delta = 3$, $\epsilon_d=0$, $N_0\sigma=200$, $V=1$ and 
	$\sqrt{D_l} \sigma_t=1/\sqrt{10}$, which corresponds to $\Gamma=0.2\pi$. }\label{fig2a}
\end{figure}
Fig.~\ref{fig2a} the top panel plots the imaginary part of $f_I$ as a
function of $\omega$ with $\bar{E}=0$ fixed.
For $\omega >0$, $f_I$ changes smoothly. And the change is smoothened
as $N_0$ increases. In the limit
$\omega \to 0^+$, $f_I$ approaches a nonzero constant, which is denoted by $f_I(\bar{E},0^+)$.
It is not difficult to determine $f_I(\bar{E},0^+)$ numerically, because
$f_I$ changes slowly as $\omega \to 0^+$ (see the curve at $N_0=7000$).
Moreover, $f_I(\omega)$ is an odd function,
hence, it must be discontinuous at $\omega=0$. This explains why we use
$f_I(\bar{E},0^+)$ instead of $f_I(\bar{E},0)$, because the latter is not well-defined. The
discontinuity at $\omega=0$ does not cause trouble in determining $f_I(\bar{E},0^+)$.
Because what we indeed calculate is $\left|f_I\right|^2$, which is continuous at $\omega=0$.

\begin{figure}[htbp]
	\centering
	\includegraphics[width=0.5\textwidth]{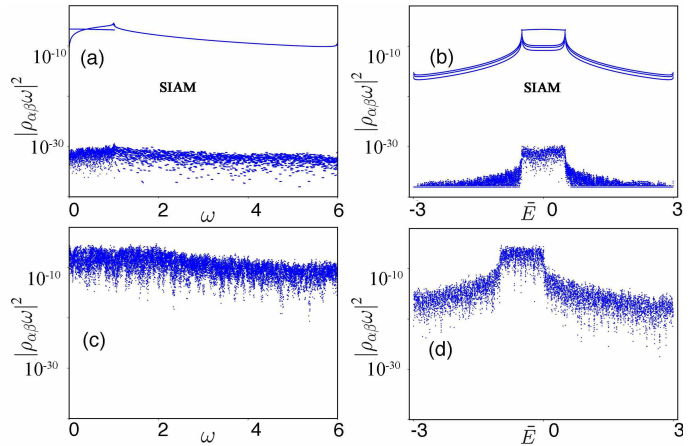}
	\caption{(Color online) (a) $\left| \rho_{\alpha,\beta}\right|^2 \omega^2$ vs $\omega$
	for a window $\bar{E}\in(-0.01,0.01)$ in the SIAM.
	(b) $\left| \rho_{\alpha,\beta}\right|^2 \omega^2$ vs $\bar{E}$
	for a window $\omega\in(0,0.01)$ in the SIAM.
	(c) $\left|\rho_{\alpha,\beta}\right|^2\omega^2$ vs $\omega$
	for a window $\bar{E}\in(-0.01,0.01)$ in our model.
	(d) $\left|\rho_{\alpha,\beta}\right|^2\omega^2$ vs $\bar{E}$
	for a window $\omega\in(0,0.01)$ in our model.
	The voltage bias is set to $V=1$.
	The other parameters are as same as those in Fig.~\ref{fig1a}.}\label{fig2aa}
\end{figure}
Next we discuss the density matrix $\rho_{\alpha,\beta}
=\bra{\alpha}\hat \rho\ket{\beta}$. Note that $\rho_{\alpha,\beta}$ is real.
Fig.~\ref{fig2aa} plots the off-diagonal elements of density matrix. To be specific, we plot
$\left|\rho_{\alpha,\beta}\right|^2 \omega^2$ vs $\omega$ for a small window
$\bar{E}\in (-0.01,0.01)$, and also $\left|\rho_{\alpha,\beta}\right|^2 \omega^2$ vs $\bar{E}$
for a small window $\omega\in(0,0.01)$. Again, the density matrix of SIAM
is plotted as a comparison. Fig.~\ref{fig2aa} makes it clear that the distribution of density
matrix has similar feature as the current matrix.
And they are qualitatively different in SIAM and our model.
We find large outliers in the density-matrix elements of SIAM. Indeed, $\left| \rho_{\alpha,\beta}\right|^2$
of SIAM are separated into two classes. A small fraction of $\left| \rho_{\alpha,\beta}\right|^2$
is much larger than the others. Conversely,
$\left| \rho_{\alpha,\beta}\right|^2$ of our model are concentrated around a smooth curve.
Fig.~\ref{fig2aa} makes it clear that the density matrix and current matrix in our model
can be expressed in a similar form. Note that Eq.~\eqref{eq:ETH}
and Eq.~\eqref{eq:NESSH} are similar to each other. In our model, $I_{\alpha,\beta}$
and $\rho_{\alpha,\beta}$ can be expressed as Eq.~\eqref{eq:ETH}
and~\eqref{eq:NESSH}, respectively. But in SIAM, neither $I_{\alpha,\beta}$
nor $\rho_{\alpha,\beta}$ can be expressed in such a form.
Indeed, ETH and NESSH usually stand simultaneously in a model, or
break down simultaneously. Their connection has been discussed in
Ref.~[\onlinecite{yang2018nonequilibrium}]

Fig.~\ref{fig2aa} presents evidence not only for the first assumption
of NESSH (Eq.~\eqref{eq:NESSH}) but also for the second assumption, i.e.
$\rho_{\alpha,\beta}$ after coarse graining diverges as $1/\omega$
as $\omega\to 0$. In the plot of $\left|\rho_{\alpha,\beta}\right|^2 \omega^2$ vs $\omega$,
a plateau is clearly seen for $\omega\in (0,2)$, indicating the $1/\omega$-divergence
of $\rho_{\alpha,\beta}$. According to Eq.~\eqref{eq:NESSH}, the
function $\left| f(\bar{E},\omega)\right|^2$ can be obtained by averaging
$\left|\rho_{\alpha,\beta}\right|^2$ over a small energy box. By using Eq.~\eqref{eq:frho},
we obtain
\begin{equation}
\rho^2(\bar{E},\omega) = D^{3}(\bar{E})\omega^2 \overline{\left| \rho_{\alpha\beta}\right|^2},
\end{equation}
where $\overline{\left| \rho_{\alpha\beta}\right|^2}$ is the average over a square
box centered at $(\bar{E},\omega)$ with the edge $0.2$. $\rho^2$
is computed in a similar way as $\left| f_I\right|^2$. Here the sign of $\rho$ can also
be chosen freely, and $\rho$ is an even function of $\omega$,
being continuous at $\omega=0$.

Fig.~\ref{fig2a} the bottom panel plots $\rho$ as a function of $\omega$
with $\bar{E}=0$ fixed. $\rho(\bar{E},\omega)$ displays a peak structure.
But it changes continuously in the vicinity of $\omega=0$. Therefore, it is not
difficult to determine the value of $\rho(\bar{E},0) $. In Fig.~\ref{fig2a}, we also
see that the characteristic functions $f_I$ and $\rho$ change with the system's size.
As $N_0$ increases, the absolute value of $f_I$ decreases, while that of
$\rho$ increases. It is worth emphasizing that the expectation of current
must be size-independent for large enough $N_0$, because
it has a well-defined thermodynamic limit.
But $f_I$ and $\rho$ need not be size-independent. Indeed, our numerics
show that $f_I$ or $\rho$ do not have thermodynamic limit.
It is not a surprise, since the density of states is also divergent as $N_0\to\infty$.
There is no paradox here, because the current comes from the product
of $f_I$ and $\rho$.

\begin{figure}[htbp]
	\centering
	\includegraphics[width=0.45\textwidth]{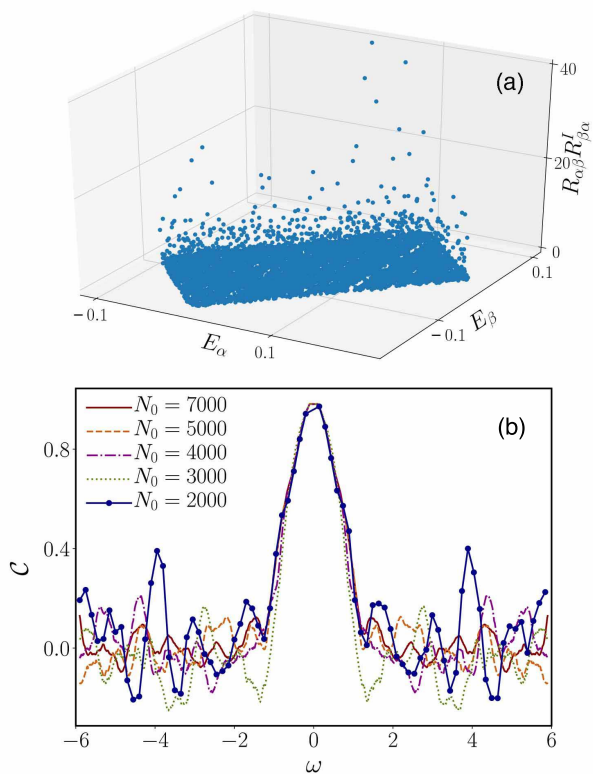}
	\caption{(Color online) (a) $R_{\alpha\beta}R^I_{\beta\alpha}$
	vs $(\bar{E},\omega)$ for the window
	$\bar{E}\in(-0.1,0.1)$ and $\omega \in (0,0.2)$. (b) $\mathcal{C}$
	as a function of $\omega$ for different $N_0$.
	The parameters are as same as those in Fig.~\ref{fig2aa}.
	}\label{fig2aaa}
\end{figure}
Finally, we discuss the random number $R_{\alpha\beta}R^I_{\beta\alpha}$.
Here $R_{\alpha\beta}$ and $R^I_{\beta\alpha}$ are real random numbers of zero mean and
unit variance, which reflect the fluctuation of $I_{\alpha,\beta}$ and $\rho_{\alpha,\beta}$,
respectively. Fig.~\ref{fig2aaa} the top panel plots
$R_{\alpha\beta}R^I_{\beta\alpha}$ vs $\bar{E}$ and $\omega$ within a small energy box.
We see that $R_{\alpha\beta}R^I_{\beta\alpha}$ distributes diversely.
Due to the correlation between $R^I_{\alpha\beta}$ and $R_{\alpha\beta}$,
the average of $R_{\alpha\beta}R^I_{\beta\alpha}$ is nonzero,
which is denoted by $\mathcal{C}$ in above.
In practice, $\mathcal{C}$ can be calculated as
\begin{equation}
\mathcal{C}(\bar{E},\omega) = \overline {R_{\alpha\beta}R^I_{\beta\alpha}},
\end{equation}
where the average is over an energy box centered at $(\bar{E},\omega)$ with
the edge chosen to $0.2$.
The hermitianity of density matrix and current matrix requires $R_{\alpha\beta}
=R_{\beta\alpha}$ and $R^I_{\alpha\beta}=R^I_{\beta\alpha}$. Therefore,
$\mathcal{C}(\bar{E},\omega)$ is an even function of $\omega$.
Fig.~\ref{fig2aaa} the bottom panel plots $\mathcal{C}$ as
a function of $\omega$ with $\bar{E}=0$ fixed.
It is clear that $\mathcal{C}$ changes smoothly as $\omega$ is in the range $(-1,1)$.
And for $\omega\in (-1,1)$, $\mathcal{C}$ is almost independent of $N_0$.
At least in the vicinity of $\omega=0$, $\mathcal{C}$ is a well-defined smooth
function of $\omega$, indicating that a numerical approach to $\mathcal{C}(\bar{E},0)$
is reliable. Since the stationary current depends only upon the value of $\mathcal{C}$
at $\omega=0$, the assumption that $\mathcal{C}$ is a smooth
function in deriving Eq.~\eqref{eq:IE} is then reliable.

So far as we can say, our numerics strongly support the assumptions
in the derivation of Eq.~\eqref{eq:IE}. We then expect Eq.~\eqref{eq:IE}
to predict the correct value of current.

\section{\label{sec:current} Current}

In this section, we discuss the NESS current.
There are two different approaches to the stationary current.
One is the statistical approach by using the formula~\eqref{eq:IE}.
The other is the ab-initio approach with respect to definition.

\begin{figure}[htbp]
	\centering
	\includegraphics[width=0.45\textwidth]{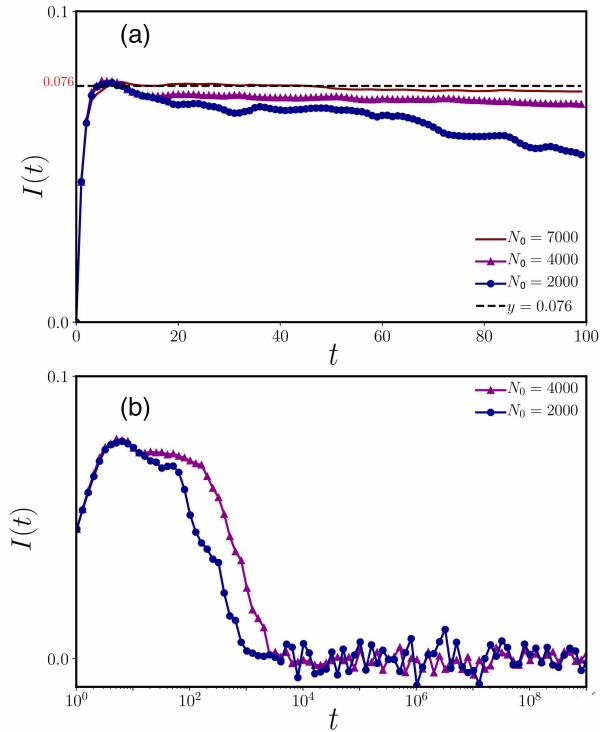}
	\caption{(Color online) The real-time dynamics of current for different
	sizes of leads. (a) $I(t)$ for $t\in (0,100)$. (b) $I(t)$ at exponentailly
	large time scales. We set $\Delta = 3$, $\epsilon_d=0$, $N_0\sigma=200$, $V=0.5$ 
	and $\sqrt{D_l}\sigma_t=1/\sqrt{10}$ (corresponding to $\Gamma\sim0.2\pi$).
	The dashed line in panel~(a) is $I=0.076$, which is obtained by the statistical approach.
	}\label{fig4}
\end{figure}
Let us start from the ab-initio approach. According to the definition of NESS current,
it is obtained by taking the limit $t\to \infty$ after $N_0\to\infty$ of
Eq.~\eqref{t-current}. While Eq.~\eqref{t-current} already tells us how to
calculate the current at arbitrary system's size and arbitrary time by using
the eigenenergies and eigenvectors of the Hamiltonian. We use this approach
to calculate $I(t)$. Fig.~\ref{fig4} plots the
real-time dynamics of current for different $N_0$. The bottom panel displays
$I(t)$ at exponentially large time scales. It is clear that $I(t)$ for arbitrary $N_0$
always decays to zero as $t\to\infty$. This is what we expect, because the
initial imbalance between two leads will be eliminated at a finite period, and
the stationary current never survives in a finite system. However, the current will first increase to
a finite value and stay there for a while before it decays.
Even for $N_0=2000$ and $N_0=4000$, we already observe a clear
plateau of $I(t)$ during the period $10<t<100$. In the top panel of Fig.~\ref{fig4},
we see that the plateau is flattened and its lifetime is enhanced
as $N_0$ increases from $2000$ to $7000$. Indeed, as $N_0=7000$,
the drop of $I$ during the period $10<t<100$ is already insignificant.
It is reasonable to conclude that the current will not drop at all as $N_0$
goes to infinity. And the height of this plateau must be
the NESS current in thermodynamic limit, that is
\begin{equation}\label{eq:Iabinitio}
I= \displaystyle\lim_{t\to\infty}
\displaystyle\lim_{N_0\to\infty} I(t).
\end{equation}

A comparison between the ab-initio result~\eqref{eq:Iabinitio}
and the statistical formula~\eqref{eq:IE} is done. In previous section,
we already explain how to obtain the characteristic functions:
$\rho(\bar{E},0)$, $f_I(\bar{E},0^+)$ and $\mathcal{C}(\bar{E},0)$.
The integral in Eq.~\eqref{eq:IE} is then carried out in a straight
forward way. The result of Eq.~\eqref{eq:IE}
is found to be $I=0.076$ for the parameters of Fig.~\ref{fig4},
which is marked as the dashed line in Fig.~\ref{fig4} the top panel.
We see that the plateau of $I(t)$ gradually approaches $0.076$
as $N_0$ increases. As $N_0=7000$, the height of plateau is approximately
equal to $0.076$. The statistical formula predicts the stationary
current to a high precision, as we expect.

\begin{figure}[htbp]
\centering
\includegraphics[width=0.45\textwidth]{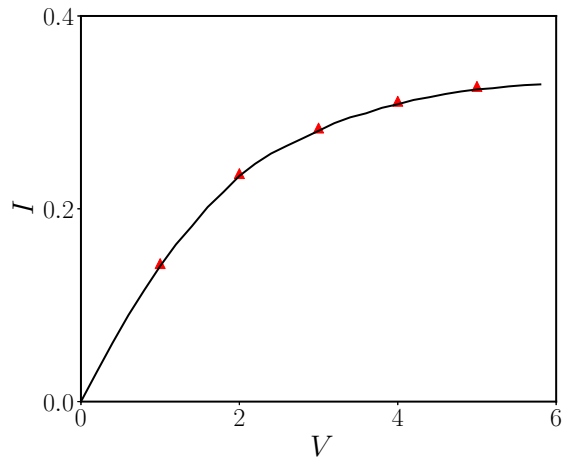}
\caption{(Color online) The current as a function of voltage bias.
	The solid line represents the ab-initio result of stationary current.
	It is obtained by taking the value of $I(t)$ at $t=40$ for the lead's size
	being $N_0=7000$. The red triangle represents the stationary current
	obtained by the statistical formula~\eqref{eq:IE}. Here we set $\Delta = 3$, $\epsilon_d=0$, 
	$N_0\sigma=200$, $N_0=7000$, and $\sqrt{D_l}\sigma_t=1/\sqrt{10}$
	(corresponding to $\Gamma\sim0.2\pi$). }\label{fig9}
\end{figure}
Fig.~\ref{fig9} plots the I-V curve. The solid line comes from the ab-initio
calculation, i.e. Eq.~\eqref{eq:Iabinitio}, while the red dots are the results of
the statistical formula~\eqref{eq:IE}. Again, the statistical formula
and the ab-initio approach predict the same stationary currents for various
voltage bias. Our results verify the current formula~\eqref{eq:IE}.
It is the first time that the statistical formula of stationary current based
on NESSH is verified in a specific model. The current is a monotonic function of voltage bias.
At small bias, $I$ increases linearly with $V$. But the current saturates at high bias.
The shape of I-V curve is reminiscent of the inverse of tangent function, while the latter
is well known to be the I-V curve of SIAM at zero temperature as the resonant level
is located at the center of transport window (see Appendix~\ref{app:SIAM} for the detail).

\begin{figure}[htbp]
	\centering
	\includegraphics[width=0.45\textwidth]{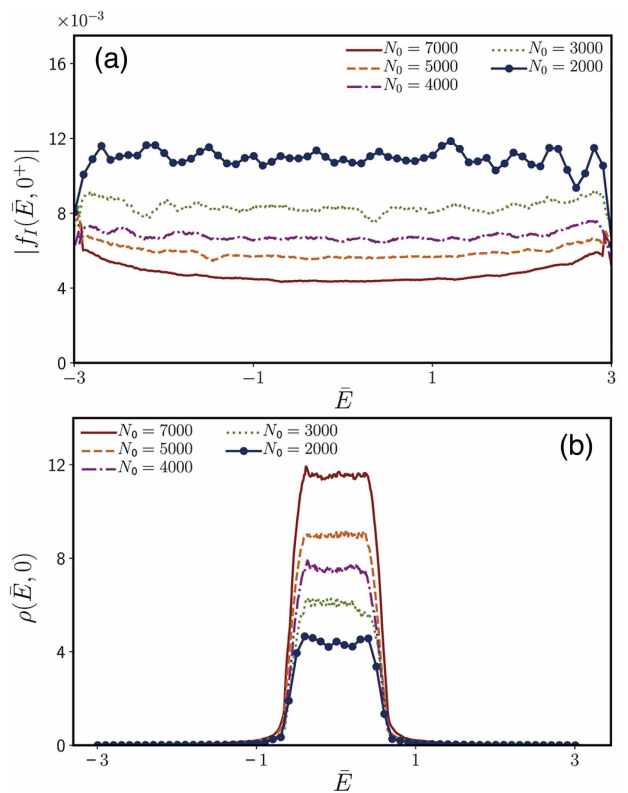}
	\caption{(Color online) (a) The function $\left|f_I(\bar{E},0^+)\right|$
	for different $N_0$. (b) The function $\rho(\bar{E},0)$
	for different $N_0$. The voltage bias is set to $V=1$.
	The other parameters are as same as
	those in Fig.~\ref{fig1}. Especially, $\sqrt{D_l}\sigma_t=\sqrt{2/5}$
	corresponds to $\Gamma = 0.8 \pi$.
	}\label{fig2}
\end{figure}
In the statistical formula, the stationary current is expressed as the integral of the product
of $\rho(\bar{E},0)$, $f_I(\bar{E},0^+)$ and $\mathcal{C}(\bar{E},0)$.
To explain the shape of I-V curve, we study these characteristic functions under different
voltage bias. Fig.~\ref{fig2}(a) displays $\left|f_I\right|$ as a function
of $\bar{E}$. Note that $f_I$ is purely imaginary,
therefore, we plot its absolute value. It is clear that $f_I$
does not change much with $\bar{E}$.
One can approximately take  $f_I(\bar{E},0^+)\approx f_I(0,0^+)$ as a constant.
On the other hand, the function $\rho$ displays a peak structure.
In the case of $\epsilon_d=0$, the peak is centered at $\bar{E}=0$ (see Fig.~\ref{fig2}(b)).
The difference between $\rho$ and $f_I$
is that $f_I$ comes from the observable operator
but $\rho$ is from the density matrix. As a consequence, $f_I$ is independent
of $V$ or the initial occupation of particles but $\rho$ depends on it.

It is worth mentioning that both $f_I$ and $\rho$ depend on the lead's size.
The magnitude of $f_I$ decreases as $N_0$ increases, while
that of $\rho$ increases. $f_I$ and $\rho$ do not have a well-defined
limit as $N_0\to \infty$ (the thermodynamic limit). Only the current has a well-defined
thermodynamic limit.

\begin{figure}[htbp]
	\centering
	\includegraphics[width=0.45\textwidth]{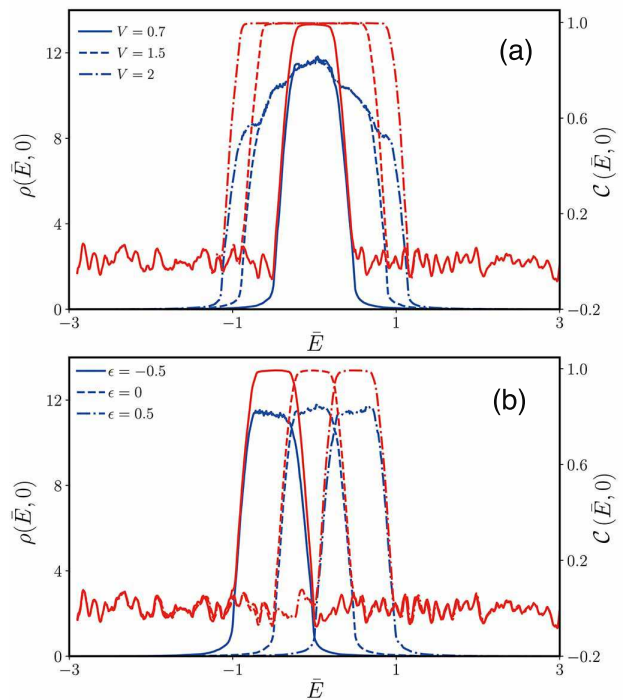}
	\caption{(Color online) (a) Three bottom lines (blue) plot the function
	$\rho(\bar{E},0)$ at different voltage bias, while three top lines (red) plot
	$\mathcal{C}(\bar{E},0)$. The functions at different $V$ are distinguished
	by different linetypes. The red and blue lines follow the same legend.
	In this panel, we set $\Delta = 3$, $\epsilon_d=0$,
	$N_o\sigma=200$, $N_0=7000$ and $\sqrt{D_l}\sigma_t=1/\sqrt{10}$
	(corresponding to $\Gamma\sim0.2\pi$). (b) We fix $V=0.7$ and
	plot the functions $\rho$ (three bottom lines) and $\mathcal{C}$
	(three top lines) at different $\epsilon_d$.
	$\epsilon_d=-0.5, 0, 0.5$ corresponds
	to the solid, dashed and dotted lines, respectively.
	}\label{fig3}
\end{figure}

The shape of $\rho(\bar{E},0)$ depends both on the initial occupation and
the position of the resonant level.
Fig.~\ref{fig3}(a) displays $\rho(\bar{E},0)$ at different voltage bias.
Here an important energy scale is the resonant level broadening $\Gamma$,
which defines the region of resonant tunneling.
When $V$ is less than $\Gamma$, $\rho$ displays a rectangular peak with a flat top,
and the width of the peak is approximately $2V$.
As $V$ increases, the peak becomes wider. But
as $V$ goes beyond $\Gamma$, the peak is not rectangular any more.
Instead, $\rho$ drops quickly to zero as $\left|\bar{E}\right|$ is larger than $\Gamma$.
The width of the peak is determined by $\Gamma$, being less than $2V$.
Fig.~\ref{fig3}(b) displays $\rho$ at different $\epsilon_d$. The center of the peak
changes with $\epsilon_d$. Indeed, the peak is approximately centered
at $\bar{E}=\epsilon_d$. But the shape of $\rho$ is indifferent to $\epsilon_d$.

The shape of $\rho(\bar{E},0)$ is reminiscent of the transmission coefficient.
For the SIAM with uniform level-spacing in leads,
the transmission coefficient is well known to be approximately
$1/\left(({E}-\epsilon_d)^2+\Gamma^2 \right)$ (see Appendix~\ref{app:SIAM}).
We see that $\rho(\bar{E},0)$ has a similar shape as
$\textbf{1}_{V} /\left((\bar{E}-\epsilon_d)^2+\Gamma^2 \right)$, where $\textbf{1}_V$
is the indicator function which equals to $1$ in the range $[-V/2,V/2]$
but zero otherwise.
As is well known, in the transport through a resonant level, the particle can tunnel from
one lead to the other if its energy is close to the resonant level,
otherwise, the particle is blocked.
It is the particles of energy around the resonant level which
contribute mainly to the current, while the other particles contribute little.
This fact is reflected by the peak structure of $\rho(\bar{E},0)$.

Fig.~\ref{fig3} also plots the correlation function $\mathcal{C}(\bar{E},0)$
at different $V$ and $\epsilon_d$. As similar as $\rho$,
the correlation function displays a peak centered at $\epsilon_d$, and the width
of the peak is approximately equal to that of $\rho$. But the peak of $\mathcal{C}$
is always rectangular with a flat top for large or small $V$. 
More important, the height of the peak is independent of $V$.
Therefore, when calculating the integral of $\rho f_I \mathcal{C}$, one can approximately
treat $\mathcal{C}$ as a constant, just like $f_I$.
How the current changes with the voltage bias is mainly determined by
the shape of $\rho(\bar{E},0)$.
Since the function $\rho(\bar{E},0)$ has a similar shape as
the transmission coefficient of SIAM,
we expect the I-V curve of our model is also similar to that of SIAM.
This is what we observe in Fig.~\ref{fig9}.
If we compare SIAM with our model,
it becomes clear that the I-V curve is robust against
replacing the uniformly-distributed levels by ones with Wigner-Dyson distribution
or replacing the constant coupling by random coupling.

\section{Summary}
\label{sec:level 4}

In the orthodox approach of studying mesoscopic transport,
the initial state is chosen to an equilibrium ensemble with
the leads being at different temperatures or chemical potentials.
The observables such as current are calculated by using an evolution
approach. After infinitely long evolution, the expectation value of observable
is believed to be equal to what is measured in experiments, even
no experimentalist has ever initialized their laboratory system in such
an equilibrium ensemble. NESSH provides a possible explanation for
the coincidence between experimental observation and theoretical
method, by stating that the microscopic properties of initial state
are unimportant in determining the long-time evolution of observable.
Especially, a current formula was derived, which connects the stationary current
to a few statistical quantities of initial state, dubbed the characteristic functions.

In this paper, we propose a specific model to check the assumptions
of NESSH and the current formula. By modeling the leads as random matrices
and the coupling between leads and resonant level as random numbers,
we find that the density matrix and current matrix fit well with the description of
NESSH and ETH, respectively. We then compute the stationary current
by the ab-initio approach and by using the current formula of NESSH.
They do agree with each other to a high precision.
According to NESSH, the stationary current is an integral
of the product of characteristic functions with respect to energy.
The characteristic functions $f_I$, $\rho$ and $\mathcal{C}$
are the coarse-grained versions of current matrix,
density matrix and their correlation, respectively.
$\rho$ as a function of energy displays a peak structure, which has
a similar shape as the transmission coefficient of SIAM.
As a consequence, the I-V curve of our model is similar to
that of SIAM with regular leads and constant coupling.

The initial density matrix and current matrix look like random matrices
in the eigenbasis of Hamiltonian. This is the key feature of our model
which is distinguished from previous models of mesoscopic transport such as SIAM.
In our model, the matrix elements have no large outliers,
but in SIAM, a small fraction of elements are dominating the others.
The concentrated distribution of off-diagonal elements around a smooth curve
is the prerequisite of defining the characteristic functions
and applying the current formula of NESSH.
In our model, the stationary current is uniquely determined by the values of
$\rho$, $f_I$ and $\mathcal{C}$ in the limit $\omega \to 0$.
This is a nonequilibrium version of memory loss.
After the system evolves into a NESS, most elements of its
initial density matrix have no contribution to physical observables. Only
the memory of a few statistical quantities is kept in the off-diagonal elements
with infinitesimal energy difference.
Our model bridge two different branches of physics - mesoscopic transport
and quantum chaos theory, and serves as a benchmark for the future study of
transport phenomena by the statistical approach based on NESSH.

\section*{acknowledgements}
This work is supported by NSF of China under Grant Nos.~11774315 and~11835011.
Pei Wang is also supported by the Junior Associates programm
of the Abdus Salam International Center for Theoretical Physics.

\appendix

\section{\label{app:SIAM} Single impurity Anderson model}

The single impurity Anderson model (SIAM) describes a resonant level
coupled to two leads. The Hamiltonian of SIAM is expressed as
\begin{equation}
\hat H = \sum_{ik} \epsilon_{ik} \hat c^\dag_{ik} \hat c_{ik} + \epsilon_d \hat d^\dag \hat d
+ g\sum_{ik} (\hat c^\dag_{ik} \hat d + h.c. ),
\end{equation}
where $\hat c^\dag_{ik}$ and $\hat c_{ik}$ are the fermionic field operators,
$i=1,2$ denotes the left and right lead, respectively, and
$k$ is the index of energy levels in each lead. In each lead,
the energy levels have a uniform distribution with the spacing between two
neighbors being a constant. We use $N_A$ to denote the number of levels
in each lead. The level position is then $\epsilon_{ik}= -\Delta+ k \displaystyle\frac{2\Delta}{N_A-1}$
with $k=0, 1, \cdots, N_A-1$, where $2\Delta$ denotes the bandwidth.
Moreover, $\hat d$ and $\hat d^\dag$ denote the field operators
of the resonant level, and $\epsilon_d$ denotes its energy. $g$ is the coupling
between the resonant level and leads. In this papers, we set $\Delta=3$,
$N_A=2000$, $\epsilon_d=0$ and $g\sqrt{N_A/(2\Delta)}= 1/\sqrt{10}$.

The current through the resonant level is usually defined to be
\begin{equation}
\begin{split}
\hat I = & \frac{1}{2}\left( \frac{d\hat N_2}{dt} -\frac{d\hat N_1}{dt}\right) \\
= & \frac{i}{2}\sum_{k,j=1}^2 (-1)^j
 g \left(\hat c^{\dagger}_{jk} \hat d- \hat d^{\dagger}\hat c_{jk}\right),
\end{split}
\end{equation}
where $\hat N_j=\sum_k \hat c^{\dagger}_{jk} \hat c_{jk}$ is the number of particles in lead $j$,
and $i$ denotes the imaginary unit. The traditional approach of calculating the stationary current
is by using the Keldysh Green's functions~\cite{Haug96}. Here an important concept is
the level broadening, which comes from the fact that
the coupling between resonant level and leads causes uncertainty in the position of resonant level.
It is defined to be $\Gamma= 2\pi D_l g^2$, where $D_l = N_A/(2\Delta)$ is
the equal density of states in both leads. With our choice of parameters,
the level broadening is $\Gamma=0.2 \pi$.
The current has a simple expression if the
bandwidth $2\Delta$ is much larger than $\Gamma$.
In this case, the transmission coefficient
is connected to the imaginary part of retarded Green's function, reading
\begin{equation}
T(E)=\frac{\Gamma^2}{\left( E-\epsilon_d \right)^2+\Gamma^2}.
\end{equation}
As the chemical potentials of the left and right leads are $V/2$ and $-V/2$, respectively,
the stationary current at zero temperature reads
\begin{equation}
\begin{split}
I = \frac{1}{2\pi} \int_{-V/2}^{V/2} dE \ T(E).
\end{split}
\end{equation}
Here we choose the atomic unit $\hbar=1$. For $\epsilon_d=0$, the I-V
curve is the inverse of tangent function.


\begin{thebibliography}{28}
\bibitem{footnote} In this paper, we do not consider the possibilities  
 of everlasting oscillation in for example a time crystal or 
 non-thermal steady state in an integrable system.
\end{thebibliography}

%

\end{document}